\documentclass[runningheads]{llncs}
\usepackage{graphicx}
\usepackage{multirow}
\usepackage{bm}
\usepackage{booktabs, makecell}
\usepackage{amsmath,amssymb}
\usepackage{hyperref}
\usepackage[capitalise]{cleveref}
\newcommand{\mcc}[1]{\multicolumn{1}{c}{#1}} 

\begin{document}
\title{{\em CRISP} - Reliable Uncertainty Estimation for Medical Image Segmentation}
%
%
\author{
Thierry Judge\inst{1} \and
Olivier Bernard\inst{3}\and 
Mihaela Porumb\inst{2} \and
Agis Chartsias\inst{2} \and 
Arian Beqiri\inst{2}\and 
Pierre-Marc Jodoin\inst{1}
}
\authorrunning{T. Judge et al.}
%
\institute{
Department of Computer Science, University of Sherbrooke, Canada \and
Ultromics Ltd., Oxford, OX4 2SU, UK \and 
University of Lyon, CREATIS, CNRS UMR5220, Inserm U1294, INSA-Lyon, University of Lyon 1, Villeurbanne, France
}
\maketitle              
\begin{abstract}
Accurate uncertainty estimation is a critical need for the medical imaging community.  A variety of methods have been proposed, all direct extensions of classification uncertainty estimations techniques. The independent pixel-wise uncertainty estimates, often based on the probabilistic interpretation of neural networks, do not take into account anatomical prior knowledge and consequently provide sub-optimal results to many segmentation tasks. For this reason, we propose {\em CRISP} a ContRastive Image Segmentation for uncertainty Prediction method.  At its core, {\em CRISP} implements a contrastive method to learn a joint latent space which encodes a distribution of valid segmentations and their corresponding images. We use this joint latent space to compare predictions to thousands of latent vectors and provide anatomically consistent uncertainty maps. Comprehensive studies performed on four medical image databases involving different modalities and organs underlines the superiority of our method compared to state-of-the-art approaches.
\keywords{Medical imaging, Segmentation, Uncertainty, Deep learning}
\end{abstract}
\section{Introduction}

Deep neural networks are the {\em de facto} solution to most segmentation, classification and clinical metric estimation. However, they provide no anatomical guarantees nor any safeguards on their predictions. Error detection and uncertainty estimation methods are therefore paramount before automatic medical image segmentation systems can be effectively deployed in clinical settings.

In this work, we present a novel uncertainty estimation method based on joint representations between images and segmentations trained with contrastive learning. Our method, {\em CRISP}
(ContRastive Image Segmentation for uncertainty Prediction), uses this representation to overcome the limitations of state-of-the-art (SOTA) methods which heavily rely on probabilistic interpretations of neural networks as is described below. 

Uncertainty is often estimated assuming a probabilistic output function by neural networks. However, directly exploiting the maximum class probability of the \textit{Softmax} or \textit{Sigmoid} usually leads to suboptimal solutions \cite{guo17_callibration}. Some improvements can be made by considering the entire output distribution through the use of entropy~\cite{settles2009active} or by using other strategies such as temperature scaling~\cite{guo17_callibration}. 

Uncertainty may also come from Bayesian neural networks, which learn a distribution over each parameter using a variational inference formalism~\cite{bayesianNN}. 
This enables weight sampling, which produces an output distribution that can model the prediction uncertainty. As Bayesian networks are difficult to train, they are often approximated by aggregating the entropy of many dropout forward runs~\cite{Gal_2015_BayesianCNN,Gal_2016_bayesNN}.  
Alternatively, a network ensemble trained with different hyper-parameters can also estimate uncertainties through differences in predictions~\cite{ensembles}.


In addition to modeling weight uncertainty, referred to as epistemic uncertainty, uncertainty in the data itself (aleatoric) can also be predicted~\cite{what_uncertainties}. However, it has been shown that these methods are less effective for segmentation~\cite{jungo_2019_assess}.

Other methods explicitly learn an uncertainty output during training. DeVries and Taylor~\cite{devries2018learning} proposed Learning Confidence Estimates (LCE) by adding a confidence output to the network. The segmentation prediction is interpolated with the ground truth according to this confidence. 
This confidence can also be learned after training by adding a confidence branch and finetuning a pre-trained network.
This enables learning the True Class Probability which is a better confidence estimate than the maximum class probability \cite{cofidnet}. 

These methods can be applied to classification and, by extension, to segmentation tasks with an uncertainty prediction at each pixel.  In theory, uncertainty maps should identify areas in which the prediction is erroneous. However, as these methods produce per-pixel uncertainties, they do not take into account higher-level medical information such as anatomical priors.  Such priors have been used in segmentation~\cite{gridnet,acnn}, but are yet to be exploited in uncertainty estimation. For instance, Painchaud et al.~\cite{Painchaud2020} remove anatomical errors by designing a latent space dedicated to the analysis and correction of erroneous cardiac shapes. However, this approach does not guarantee that the corrected shape matches the input image.



To this end, we propose {\em CRISP}, a method which does not take into account the probabilistic nature of neural networks, but rather uses a joint latent representation of anatomical shapes and their associated image. This paper will describe the {\em CRISP} method and propose a rigorous evaluation comparing {\em CRISP} to SOTA methods using four datasets. 


\section{{\em CRISP}}
\begin{figure*}[tp]
\centering
\includegraphics[width=0.85\textwidth]{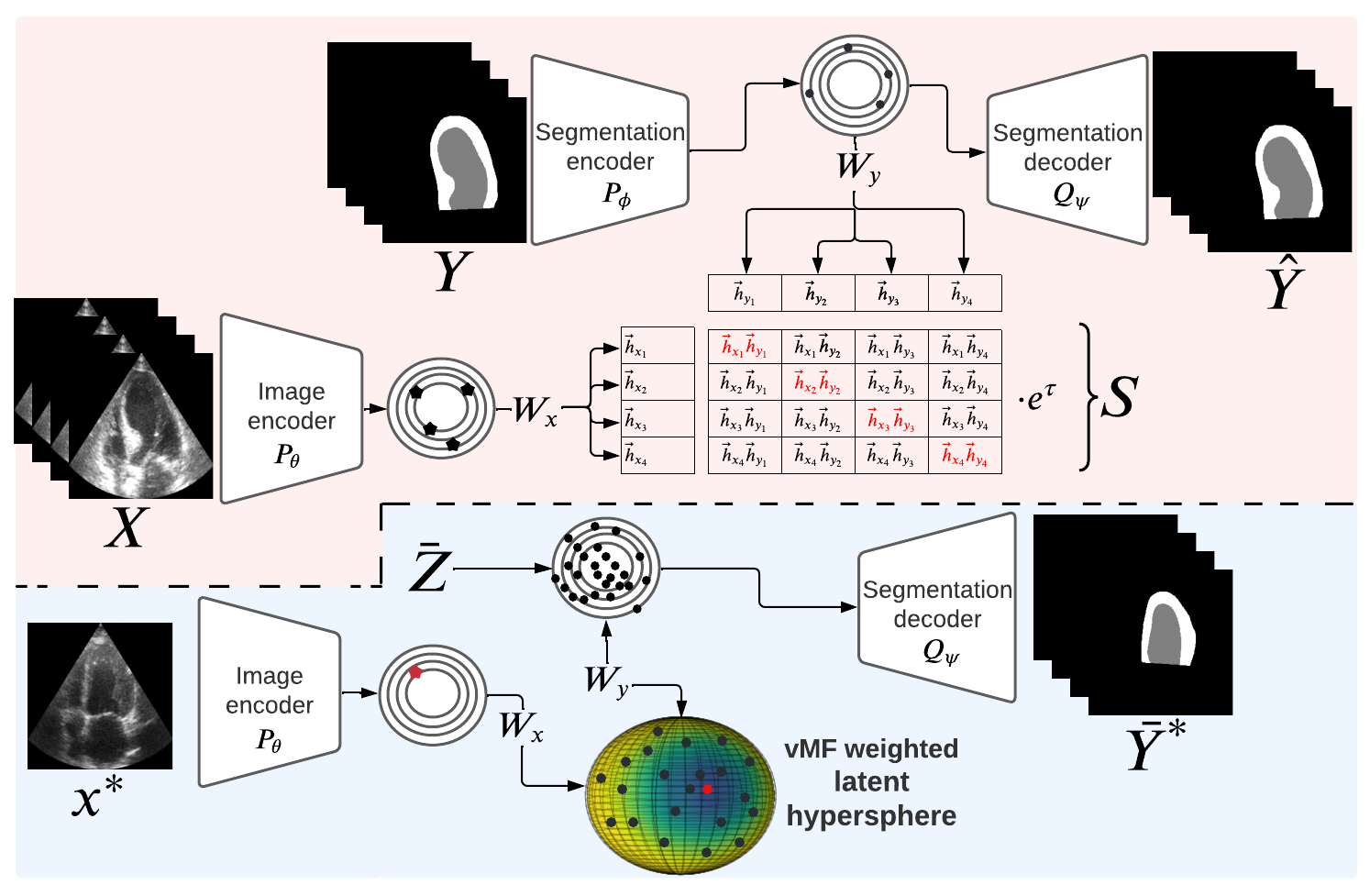}
\caption{Schematic representation of our method. Top depicts the training phase and bottom illustrate the uncertainty estimation on an input-prediction pair $(x^*, y^*)$.}
\label{fig:method}
\end{figure*}


The overarching objective of our method is to learn a joint latent space, in which the latent vector of an input image  lies in the vicinity of its corresponding segmentation map's latent vector in a similar fashion as the ``CLIP'' method does for images and text~\cite{clip}.  As such, a test image $x$ whose latent vector does not lie close to that of its segmentation map $y$ is an indication of a potentially erroneous segmentation. 
Further details are given below.

\noindent \textbf{Training.}
As shown in \cref{fig:method}, at train time, {\em CRISP} is composed of two encoders: the image encoder $P_\theta$ and the segmentation encoder $P_\phi$.  They respectively encode an image 
$x_i$ and its associated segmentation groundtruth 
$y_i$ into latent vectors $\vec{z}_{x_i}\in \Re^{D_x}$ and $\vec{z}_{y_i}\in \Re^{D_y}$ $\forall i$.  Two weight matrices $W_x \in \Re^{D_h \times D_x}$ and $W_y \in \Re^{D_h \times D_y}$ linearly project the latent vectors into a joint $D_h-$dimensional latent space where samples are normalized and thus projected onto a hyper-sphere. As such, the image latent vector $\vec{z}_{x_i}$ is projected onto a vector $\vec{h}_{x_i} = \frac{W_x \cdot \vec{z}_{x_i}}{|| W_x \cdot \vec{z}_{x_i} ||}$ and similarly for  $\vec{z}_{y_i}$.  A successful training should lead to a joint representation for which  $\vec{h}_{x_i} \approx \vec{h}_{y_i}$.

 During training, images and groundtruth maps are combined into  batches of $B$ elements, $\bm{X}=[x_1 x_2...x_B] \in \Re^{\scriptscriptstyle B\!\times\! C\!\times\! H\!\times\! W}$  and $\bm{Y}=[y_1y_2...y_B] \in \{0,1\}^{\scriptscriptstyle B\!\times\! K\!\times\! H\!\times\! W}$ for images with $C$ channels and $K$ segmentation classes.
As mentioned before, these batches are encoded by $P_\theta$ and $P_\phi$ into sets of latent vectors $Z_X$ and $Z_Y$ and then projected and normalized into sets of joint latent vectors $H_X$ and  $H_Y$. 

At this point, a set of $2\times B$ samples lie on the surface of a unit hyper-sphere of the joint latent space.  Much like CLIP~\cite{clip}, the pair-wise distance between these joint latent vectors is computed with a cosine similarity that we scale by a learned temperature factor $\tau$ to control the scale of the logits. This computation is done by taking a weighted product between $H_X$ and $H_Y$ which leads to the following square matrix: $S = (H_X \cdot H_Y^T) e^\tau \in \Re^{B\times B}$.  As shown in \cref{fig:method},  the diagonal of $S$ corresponds to the cosine similarity of the latent image vectors with their corresponding latent groundtruth vector while the off-diagonal elements are cosine similarity of unrelated vectors.  

The goal during training is to push $S$ towards an identity matrix, such that the latent vectors $\vec{h}_{x_i}$ and $\vec{h}_{y_j}$ lie on the same spot in the joint latent space when $i=j$ and are orthogonal when $i\not = j$.  This would lead to similarities close to 1 on the diagonal and close to 0 outside of it.
To enforce this, a cross-entropy loss on the rows and columns of $S$ is used as a contrastive loss~\cite{clip}, 

\begin{equation}
    \mathcal{L}_{cont} = - \frac{1}{2}\left (\frac{1}{B}\sum_{i=1}^B\sum_{j=1}^B I_{ij} \log{S_{ij}} + \frac{1}{B}\sum_{i=1}^B\sum_{j=1}^B I_{ji} \log{S_{ji}} \right).
\end{equation}

{\em CRISP} also has a segmentation decoder $Q_\psi$ to reconstruct segmentation latent vectors, a critical feature for estimating uncertainty.  This decoder is trained with a reconstruction loss $\mathcal{L}_{rec}$ which is a weighted sum of the Dice coefficient and cross-entropy loss. 
The model is trained end-to-end to minimize $\mathcal{L} = \mathcal{L}_{cont} + \mathcal{L}_{rec}$.

\noindent{\bf Uncertainty prediction.} 
Once training is over, the groundtruth segmentation maps $\bm{Y}$ are projected one last time into the Z and H latent spaces.  This leads to a set of $N$ latent vectors $\bar{Z} \in \Re^{N \times D_y}$ and $\bar{H} \in \Re^{N \times D_h}$ which can be seen as latent anatomical prior distributions that will be used to estimate uncertainty.

Now let $x^*$ be a non-training image and $y^*$ its associated segmentation map computed with a predetermined segmentation method (be it a deep neural network or not).  To estimate an uncertainty map, $x^*$ is projected into the joint latent space to get its latent vector $\vec h_{x^*} \in \Re^{D_h}$.  We then compute a weighted dot product between $\vec h_{x^*}$ and each row of $\bar{H}$ to get $\bar S\in \Re^N$, a vector of similarity measures between $\vec h_{x^*}$ and every groundtruth latent vector.  Interestingly enough, the way {\em CRISP} was trained makes $\bar S$ a similarity vector highlighting how each groundtruth map fits the input image $x^*$.   



Then, the $M$ samples of $\bar{Z}$ with the highest values in $\vec{S}$ are selected. These samples are decoded to obtain $\bar{Y}^*$, {\em i.e.} various anatomically valid segmentation maps whose shapes are all roughly aligned on $x^*$.  To obtain an uncertainty map, we compare these samples to the initial prediction $y^*$. We compute the average of the pixel-wise difference between $y^*$ and $\bar{Y}^*$ to obtain an uncertainty map $U$. 

\begin{equation}
U = \frac{1}{M}\sum_{i=1}^{M} w_i(\bar{y}^*_i - y^*)
\end{equation}

As not all samples equally correspond to $x^*$, we add a coefficient $w_i$ which corresponds to how close a groundtruth map $y_i$ is from $x^*$.  Since the joint latent space is a unit hyper-sphere, we use a  \textit{von Mises-Fisher distribution} (vMF) \cite{Directionalstats} centered on $\vec{h}_{x^*}$ as a kernel to weigh its distance to  $\vec{h}_{y_i}$. We use Taylor's method~\cite{TAYLOR_kernal_bandwith} to define the kernel bandwidth $b$ (more details are available in the supplementary materials). We define the kernel as:
%

\begin{equation}
    w_i = e^{\frac{1}{b} \vec{h_i}^T \vec{h}_{x^*}} / e^{\frac{1}{b} \vec{h}_{x^*}^T \vec{h}_{x^*}} = e^{\frac{1}{b} (\vec{h_i}^T \vec{h}_{x^*} - 1)}.
\end{equation}

\section{Experimental setup}

\subsection{Uncertainty metrics}



\noindent \textbf{Correlation.} Correlation is a straightforward method for evaluating the quality of uncertainty estimates for a full dataset. The absolute value of the Pearson correlation score is computed between the sample uncertainty and the Dice score. In this paper, sample uncertainty is obtained by dividing the sum of the uncertainty for all pixels by the number of foreground pixels. Ideally, the higher the Dice is, the lower  the sample uncertainty should be. Therefore, higher correlation values indicate more representative uncertainty maps.

\noindent \textbf{Calibration.} A classifier is calibrated if its confidence is equal to the probability of being correct. 
Calibration is expressed with Expected Calibration Error (ECE) computed by splitting all $n$ samples into $m$ bins and computing the mean difference between the accuracy and average 
confidence for each bin.  Please refer to the following paper for more details~\cite{ece}.
%


\noindent \textbf{Uncertainty-error mutual information.} Previous studies have computed Uncertainty-error overlap obtaining the Dice score between the thresholded uncertainty map and a pixel-wise error map between the prediction and the ground-truth segmentation map~\cite{jungo_2019_assess}. As the uncertainty error overlap requires the uncertainty map to be thresholded, much of the uncertainty information is lost. We therefore propose computing the mutual information between the raw uncertainty map and the pixel-wise error map. We report the average over the test set weighted by the sum of erroneous pixels in the image. 


\subsection{Data}

\noindent \textbf{CAMUS.} The CAMUS dataset~\cite{Leclerc19} consists of cardiac ultrasound clinical exams performed on 500 patients.  
Each exam contains the 2D apical four-chamber (A4C) and two-chamber view (A2C) sequences. 
Manual delineation of the endocardium and epicardium borders of the left ventricle (LV) and atrium were made by a cardiologist for the end-diastolic (ED) and end-systolic (ES) frames. The dataset is split into training, validation and testing sets of 400, 50 and 50 patients respectively.

\noindent \textbf{HMC-QU.} The HMC-QU dataset~\cite{hmc_qu} is composed of 162 A4C and 130 A2C view recordings. 
93 A4C and 68 A2C sequences correspond to patients with scarring from myocardial infarction. The myocardium (MYO) of 109 A4C (72 with myocardial infarction/37 without) recordings was manually labeled for the full cardiac cycle. These sequences were split into training, validation and testing sets of 72, 9 and 28 patients.


\noindent \textbf{Shenzen.} The Shenzen dataset~\cite{Jaeger2014} is a lung X-ray dataset acquired for pulmonary tuberculosis detection. The dataset contains 566 postero-anterior chest radiographs and corresponding manually segmented masks to identify the lungs. The dataset was split into training and validation sets of 394 and 172 patients.

\noindent \textbf{JSRT.} We use the Japanese Society of Radiological Technology (JSRT)~\cite{jstr} lung dataset which contains images and segmentation maps for 154 radiographs with lung nodules, and corresponding segmentation masks.

\subsection{Implementation details}

{\em CRISP} was compared to several SOTA methods mentioned before.  To make comparison fair, every method use the same segmentation network (an Enet~\cite{Paszke_2016_enet} in our case). All methods were trained with a batch size of 32 and the Adam optimizer~\cite{adam} with a learning rate of 0.001 and weight decay of 1e-4. We added early stopping and selected the weights with the lowest validation loss. The \textbf{Entropy} method was tested using the baseline network. We tested \textbf{MC Dropout} by increasing the baseline dropout value from 10\% to 25\% and 50\% (we report best results with respect to the Dice score) and computing the average of 10 forward passes.  For \textbf{LCE}, we duplicated the last bottleneck of the Enet to output confidence. The \textbf{Confidnet} method was trained on the baseline Enet pre-trained network. The full decoder was duplicated to predict the True Class Probability. 
For methods or metrics that require converting pixel-wise confidence (c) to uncertainty (u), we define the relationship between the two as $u = 1 - c$ as all methods produce values in the range $[0,1]$.

To highlight some limitations of SOTA methods, we also added a na\"ive method for computing uncertainty which we referred to as \textbf{Edge}. The uncertainty map for \textit{Edge} amounts to a trivial edge detector applied to the predicted segmentation map.  The resulting borders have a width of 5 pixels. 

As our \textbf{CRISP} method can be used to evaluate any image-segmentation pair, regardless of the segmentation method, we tested it on all the segmentation methods that produce different results (baseline, MC Dropout, LCE). This allows for a more robust evaluation as the evaluation of uncertainty metrics is directly influenced by the quality of the segmentation maps \cite{jungo_2019_assess}. The value of $M$ was determined empirically and kept proportional to the size of $\bar{Z}$. It can be noted, that the vMF weighting in the latent space attenuates the influence of $M$.

\subsection{Experimental setup}

We report results on both binary and multi-class segmentation tasks. As our datasets are relatively large and homogeneous, Dice scores are consistently high. This can skew results as methods can simply predict uncertainty around the prediction edges. Thus, as mentioned below, we tested on different datasets or simulated domain shift through data augmentation. 

\begin{figure*}[tp]
\centering
\includegraphics[width=0.8\textwidth]{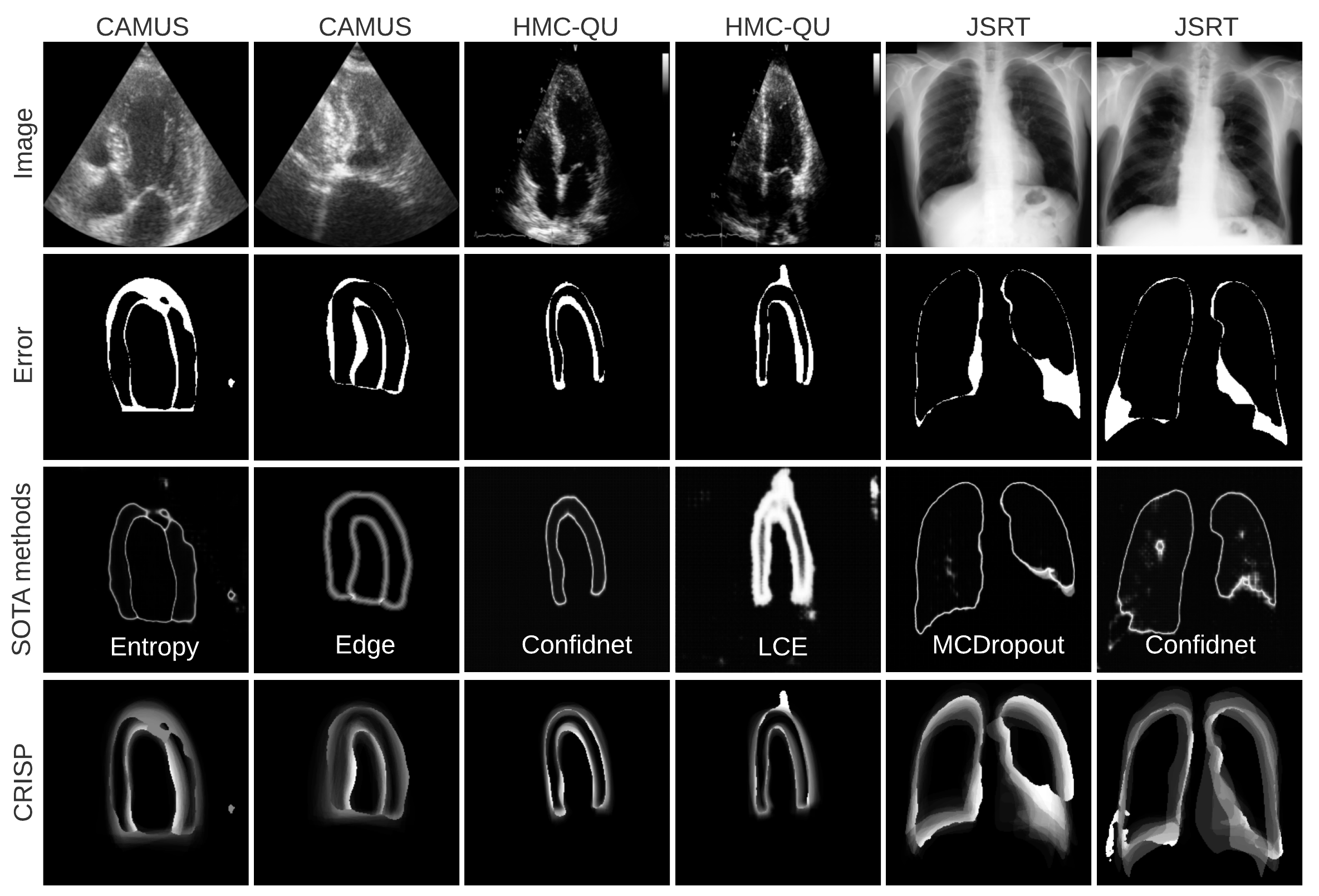}
\caption{From top to bottom: raw images, corresponding error maps, uncertainty estimation of SOTA methods and CRISP uncertainty. White indicates erroneous pixels in the error maps [row 2] and high uncertainty in the uncertainty maps [rows 3 and 4].}
\label{fig:samples}
\end{figure*}

Tests were conducted on the CAMUS dataset for LV and MYO segmentation. 
We simulated a domain shift by adding brightness and contrast augmentations (factor=0.2) and Gaussian noise ($\sigma^2=0.0001$) with probability of 0.5 for all test images. 
We used the 1800 samples from the training and validation sets to make up the $\bar{Z}$ set and used $M=50$ samples to compute the uncertainty map.

We also tested all methods trained on the CAMUS dataset on the HMC-QU dataset for myocardium segmentation. We added brightness and contrast augmentations (factor=0.2) and RandomGamma (0.95 to 1.05) augmentations during training and normalized the HMC-QU samples using the mean and variance of the CAMUS dataset. We used the A4C samples from the CAMUS dataset (along with interpolated samples between ES and ED instants) to create the set of latent vectors $\bar{Z}$. This corresponds to 8976 samples, of which $M=150$ were selected to compute the uncertainty map. 

Finally, we tested our method on a different modality and organ by using the lung X-ray dataset. We trained all the methods on the Shenzen dataset and tested on the JSRT dataset. We normalized JSRT samples with the mean and variance of the Shenzen dataset. We used the 566 samples from the Shenzen dataset to form the $\bar{Z}$ set and used $M=25$ samples to compute the uncertainty.

\begin{table}[tp]
    \centering
    \small
    \begin{tabular*}{\textwidth}{l @{\extracolsep{\fill}} ccccccccc}
    \toprule
    Training data & \multicolumn{3}{c}{CAMUS} & \multicolumn{3}{c}{CAMUS}   & \multicolumn{3}{c}{Shenzen} \\
    Testing data & \multicolumn{3}{c}{CAMUS} & \multicolumn{3}{c}{HMC-QU}   & \multicolumn{3}{c}{JSRT} \\
     \cmidrule(lr){1-1} \cmidrule(lr){2-4}  \cmidrule(lr){5-7}  \cmidrule(lr){8-10} 
    Method & \mcc{Corr. $\uparrow$} & \mcc{ECE $\downarrow$} & \mcc{MI $\uparrow$}   & \mcc{Corr. $\uparrow$} & \mcc{ECE $\downarrow$} & \mcc{MI $\uparrow$}   & \mcc{Corr. $\uparrow$} & \mcc{ECE $\downarrow$} & \mcc{MI $\uparrow$}  \\
    \midrule\midrule
    Entropy                     &          0.66 &                     0.12  &            0.02  &            0.34 &                      0.27 &            0.02    & \textbf{0.89} &                      0.08 &            0.02 \\
    Edge                        &          0.64 &            \textbf{0.06}  &            0.05  &            0.15 &             \textbf{0.08} &   \textbf{0.06}    &          0.81 &             \textbf{0.05} &            0.03 \\
    ConfidNet                   &          0.34 &                     0.08  &            0.04  &            0.36 &                      0.17 &            0.04    &          0.69 &                      0.09 &            0.01 \\
    {\em CRISP}                 & \textbf{0.71} &                     0.09  &   \textbf{0.20}  &   \textbf{0.41} &                      0.14 &   \textbf{0.06}    &          0.83 &                      0.19 &   \textbf{0.11} \\
    \midrule
    McDropout                   &          0.67 &                     0.13  &            0.03  &            0.26 &                      0.26 &            0.02    & \textbf{0.82} &              \textbf{0.06} &            0.03 \\
    {\em CRISP}-MC              & \textbf{0.78} &            \textbf{0.11}  &   \textbf{0.26}  &   \textbf{0.29} &             \textbf{0.14} &   \textbf{0.06}    & \textbf{0.82} &                      0.21 &    \textbf{0.08} \\
    \midrule
    LCE                         &          0.58 &                     0.44  &            0.08  &   \textbf{0.35} &                      0.37 &   \textbf{0.07}    & \textbf{0.87} &                      0.37 &            0.06 \\
    {\em CRISP}-LCE             & \textbf{0.59} &            \textbf{0.08}  &   \textbf{0.15}  &            0.34 &             \textbf{0.13} &   \textbf{0.07}    &          0.85 &              \textbf{0.18} &    \textbf{0.11} \\
    \bottomrule &
    \end{tabular*}
    \caption{Uncertainty estimation results (average over 3 random seeds) for different methods. Bold values indicate best results.}
    \label{tab:resutls}
\end{table}

\section{Results}

Uncertainty maps are presented in \cref{fig:samples} for samples on 3 datasets.  As can be seen, {\em Entropy, ConfidNet}, and {\em MCDropout} have a tendency to work as an edge detector, much like the naive {\em Edge} method.  As seen in \Cref{tab:resutls}, different methods perform to different degrees on each of the datasets. However, CRISP is consistently the best or competitive for all datasets for the correlation and MI metrics.  ECE results for {\em CRISP} are also competitive but not the best. Interestingly, the trivial {\em Edge} method often reports the best ECE results. This is probably due to the fact that errors are more likely to occur near the prediction boundary and the probability of error decreases with distance. These results might encourage the community to reconsider the value of ECE for specific types of segmentation tasks. 

\begin{figure*}[tp]
\centering
\includegraphics[width=0.7\textwidth]{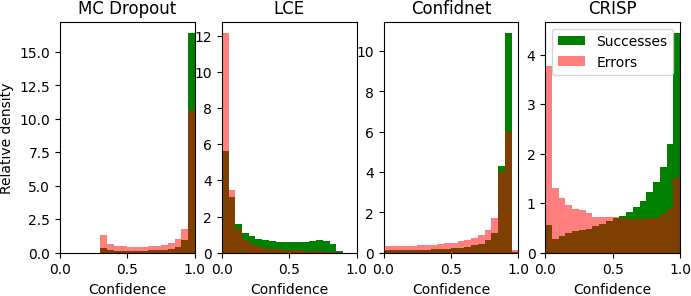}
\caption{Histograms of well classified pixels (Successes) and mis-classified pixels (Errors) for different methods on the HMC-QU dataset.}
\label{fig:hist}
\end{figure*}

\cref{fig:hist} shows the distribution of pixel confidence according to the well-classified and misclassified pixels. This figure allows for a better understanding of the different shortcomings of each method. It clearly shows that both MC Dropout and \textit{Confidnet} methods produce over-confident results. On the other hand, LCE appears to produce slightly under-confident predictions which explains the higher mutual information value. Finally, {\em CRISP} is the only method that can clearly separate certain and uncertain pixels. These results are consistent with what is observed in \cref{fig:samples} as both MC Dropout and \textit{Confidnet} produce very thin uncertainty and LCE predicts large areas of uncertainty around the border. Only {\em CRISP} produces varying degrees of uncertainty according to the error. 

It is apparent that there is a slight decrease in performance for {\em CRISP}  on the JSRT dataset. This is most likely caused by the fact that the latent space is not densely populated during uncertainty estimation. Indeed, the 566 samples in $\bar{Z}$ might not be enough to produce optimal uncertainty maps. This is apparent in \cref{fig:samples} where the uncertainty maps for the JSRT samples are less smooth than the other datasets that have more latent vectors. Different techniques such as data augmentation or latent space rejection sampling~\cite{Painchaud2020} are plausible solutions. 

\section{Discussion and conclusion}

While empirical results indicate that all methods perform to a certain degree, qualitative results in \cref{fig:samples} show that most SOTA methods predict uncertainty around the prediction edges. While this may constitute a viable uncertainty prediction when the predicted segmentation map is close to the groundtruth, these uncertainty estimates are useless for samples with large errors. Whereas in other datasets and modalities, the uncertainty represents the probability of a structure being in an image and at a given position, lung X-Ray and cardiac ultrasound structures are always present and are of regular shape and position. This makes the task of learning uncertainty during training challenging as few images in the training set produce meaningful errors. Compared to other approaches, {\em CRISP}  leverages the information contained in the dataset to a greater degree and accurately predicts uncertainty in even the worst predictions.




To conclude, we have presented a method to  identify uncertainty in segmentation by exploiting a joint latent space trained using contrastive learning. We have shown that SOTA methods produce sub-optimal results due to the lack of variability in segmentation quality during training when segmenting regular shapes.  We also highlighted this with the na\"ive {\em Edge} method. However, due to its reliance on anatomical priors, {\em CRISP} can identify uncertainty in a wide range of segmentation predictions. 

\clearpage
\bibliographystyle{splncs04}
\bibliography{mybibliography}

\begin{thebibliography}{10}
\providecommand{\url}[1]{\texttt{#1}}
\providecommand{\urlprefix}{URL }
\providecommand{\doi}[1]{https://doi.org/#1}

\bibitem{cofidnet}
Corbi\`{e}re, C., Thome, N., Bar-Hen, A., Cord, M., P\'{e}rez, P.: Addressing
  failure prediction by learning model confidence. In: Advances in Neural
  Information Processing Systems 32, pp. 2902--2913. Curran Associates, Inc.
  (2019)

\bibitem{hmc_qu}
Degerli, A., Zabihi, M., Kiranyaz, S., Hamid, T., Mazhar, R., Hamila, R.,
  Gabbouj, M.: Early detection of myocardial infarction in low-quality
  echocardiography. IEEE Access  \textbf{9},  34442--34453 (2021)

\bibitem{devries2018learning}
DeVries, T., Taylor, G.W.: Learning confidence for out-of-distribution
  detection in neural networks. arXiv preprint arXiv:1802.04865  (2018)

\bibitem{Gal_2015_BayesianCNN}
Gal, Y., Ghahramani, Z.: Bayesian convolutional neural networks with bernoulli
  approximate variational inference. ArXiv  \textbf{abs/1506.02158} (2015)

\bibitem{Gal_2016_bayesNN}
Gal, Y., Ghahramani, Z.: Dropout as a bayesian approximation: Representing
  model uncertainty in deep learning. In: Proceedings of the 33rd International
  Conference on International Conference on Machine Learning - Volume 48. p.
  1050–1059. ICML’16, JMLR.org (2016)

\bibitem{guo17_callibration}
Guo, C., Pleiss, G., Sun, Y., Weinberger, K.Q.: On calibration of modern neural
  networks. In: Precup, D., Teh, Y.W. (eds.) Proceedings of the 34th
  International Conference on Machine Learning. Proceedings of Machine Learning
  Research, vol.~70, pp. 1321--1330. PMLR (06--11 Aug 2017)

\bibitem{Jaeger2014}
Jaeger, S., Candemir, S., Antani, S., Wáng, Y.X.J., Lu, P.X., Thoma, G.: Two
  public chest x-ray datasets for computer-aided screening of pulmonary
  diseases. Quantitative Imaging in Medicine and Surgery  \textbf{4}, ~475
  (2014)

\bibitem{jungo_2019_assess}
Jungo, A., Reyes, M.: Assessing reliability and challenges of uncertainty
  estimations for medical image segmentation. In: Shen, D., Liu, T., Peters,
  T.M., Staib, L.H., Essert, C., Zhou, S., Yap, P.T., Khan, A. (eds.) Medical
  Image Computing and Computer Assisted Intervention -- MICCAI 2019. pp.
  48--56. Springer International Publishing, Cham (2019)

\bibitem{bayesianNN}
Kendall, A., Gal, Y.: What uncertainties do we need in bayesian deep learning
  for computer vision? In: Guyon, I., Luxburg, U.V., Bengio, S., Wallach, H.,
  Fergus, R., Vishwanathan, S., Garnett, R. (eds.) Advances in Neural
  Information Processing Systems. vol.~30, pp. 5574--5584. Curran Associates,
  Inc. (2017)

\bibitem{what_uncertainties}
Kendall, A., Gal, Y.: What uncertainties do we need in bayesian deep learning
  for computer vision? In: Guyon, I., Luxburg, U.V., Bengio, S., Wallach, H.,
  Fergus, R., Vishwanathan, S., Garnett, R. (eds.) Advances in Neural
  Information Processing Systems. vol.~30. Curran Associates, Inc. (2017)

\bibitem{adam}
Kingma, D.P., Ba, J.: Adam: {A} method for stochastic optimization. In: Bengio,
  Y., LeCun, Y. (eds.) 3rd International Conference on Learning
  Representations, {ICLR} 2015, San Diego, CA, USA, May 7-9, 2015, Conference
  Track Proceedings (2015)

\bibitem{ensembles}
Lakshminarayanan, B., Pritzel, A., Blundell, C.: Simple and scalable predictive
  uncertainty estimation using deep ensembles. In: Guyon, I., Luxburg, U.V.,
  Bengio, S., Wallach, H., Fergus, R., Vishwanathan, S., Garnett, R. (eds.)
  Advances in Neural Information Processing Systems. vol.~30. Curran
  Associates, Inc. (2017)

\bibitem{Leclerc19}
Leclerc, S., Smistad, E., Pedrosa, J., Østvik, A., Cervenansky, F., Espinosa,
  F., Espeland, T., Berg, E.A.R., Jodoin, P.M., Grenier, T., Lartizien, C.,
  D’hooge, J., Lovstakken, L., Bernard, O.: Deep learning for segmentation
  using an open large-scale dataset in 2d echocardiography. IEEE Transactions
  on Medical Imaging  \textbf{38}(9),  2198--2210 (2019)

\bibitem{Directionalstats}
Mardia, K.V., Jupp, P.E.: Directional Statistics. Wiley (1999)

\bibitem{acnn}
Oktay, O., Ferrante, E., Kamnitsas, K., Heinrich, M., Bai, W., Caballero, J.,
  Cook, S.A., de~Marvao, A., Dawes, T., O‘Regan, D.P., Kainz, B., Glocker,
  B., Rueckert, D.: Anatomically constrained neural networks (acnns):
  Application to cardiac image enhancement and segmentation. IEEE Transactions
  on Medical Imaging  \textbf{37}(2),  384--395 (2018)

\bibitem{Painchaud2020}
Painchaud, N., Skandarani, Y., Judge, T., Bernard, O., Lalande, A., Jodoin,
  P.M.: Cardiac segmentation with strong anatomical guarantees. IEEE
  Transactions on Medical Imaging  \textbf{39}(11),  3703--3713 (2020)

\bibitem{ece}
Pakdaman~Naeini, M., Cooper, G., Hauskrecht, M.: Obtaining well calibrated
  probabilities using bayesian binning. Proceedings of the AAAI Conference on
  Artificial Intelligence  \textbf{29}(1) (Feb 2015)

\bibitem{Paszke_2016_enet}
Paszke, A., Chaurasia, A., Kim, S., Culurciello, E.: Enet: {A} deep neural
  network architecture for real-time semantic segmentation. CoRR
  \textbf{abs/1606.02147} (2016)

\bibitem{clip}
Radford, A., Kim, J.W., Hallacy, C., Ramesh, A., Goh, G., Agarwal, S., Sastry,
  G., Askell, A., Mishkin, P., Clark, J., Krueger, G., Sutskever, I.: Learning
  transferable visual models from natural language supervision. CoRR
  \textbf{abs/2103.00020} (2021)

\bibitem{settles2009active}
Settles, B.: Active learning literature survey. Computer Sciences Technical
  Report~1648, University of Wisconsin--Madison (2009)

\bibitem{jstr}
Shiraishi, J., et~al.: Development of a digital image database for chest
  radiographs with and without a lung nodule: receiver operating characteristic
  analysis of radiologists' detection of pulmonary nodules. American Journal of
  Roentgenology  \textbf{174},  71--74 (2000)

\bibitem{TAYLOR_kernal_bandwith}
Taylor, C.C.: Automatic bandwidth selection for circular density estimation.
  Computational Statistics and Data Analysis  \textbf{52}(7),  3493--3500
  (2008)

\bibitem{gridnet}
Zotti, C., Humbert, O., Lalande, A., Jodoin, P.M.: Gridnet with automatic shape
  prior registration for automatic mri cardiac segmentation. MICCAI - ACDC
  Challenge  (2017)

\end{thebibliography}


\begin{thebibliography}{1}
\providecommand{\url}[1]{\texttt{#1}}
\providecommand{\urlprefix}{URL }
\providecommand{\doi}[1]{https://doi.org/#1}

\bibitem{kappa_estimate}
Banerjee, A., Dhillon, I.S., Ghosh, J., Sra, S.: Clustering on the unit
  hypersphere using von mises-fisher distributions. Journal of Machine Learning
  Research  \textbf{6}(46),  1345--1382 (2005)

\bibitem{TAYLOR_kernal_bandwith}
Taylor, C.C.: Automatic bandwidth selection for circular density estimation.
  Computational Statistics and Data Analysis  \textbf{52}(7),  3493--3500
  (2008)

\end{thebibliography}

\end{document}


\title{CRISP - Reliable Uncertainty Estimation for Medical Image Segmentation \\ \small{Supplementary materials\vspace{-1cm}}}
\titlerunning{CRISP- Supplementary materials}
%
\author{
}
%
%
\institute{
}
\maketitle              

\vspace{-0.3cm}
\section{Ablation study}
\vspace{-0.5cm}
\begin{table}[h!]
    \centering
    \small
    \begin{tabular*}{\textwidth}{l @{\extracolsep{\fill}} ccccccccc}
    \toprule
    Training data & \multicolumn{3}{c}{CAMUS} & \multicolumn{3}{c}{CAMUS}   & \multicolumn{3}{c}{Shenzen} \\
    Testing data & \multicolumn{3}{c}{CAMUS} & \multicolumn{3}{c}{HMC-QU}   & \multicolumn{3}{c}{JSRT} \\
     \cmidrule(lr){1-1} \cmidrule(lr){2-4}  \cmidrule(lr){5-7}  \cmidrule(lr){8-10} 
    Method & \mcc{M} & \mcc{Corr. $\uparrow$}  & \mcc{MI $\uparrow$}   & \mcc{M} & \mcc{Corr. $\uparrow$}  & \mcc{MI $\uparrow$} &  \mcc{M} & \mcc{Corr. $\uparrow$}  & \mcc{MI $\uparrow$}  \\
    \midrule\midrule

    {\em CRISP} &       25  &             0.72 &             0.20            &  50   &   0.41 &     0.06        & 5   &             0.83 &             0.10 \\
    {\em CRISP} &       50  &             0.71 &             0.20            &  100  &   0.41 &     0.06        & 10  &             0.84 &             0.11 \\
    {\em CRISP} &       100 &             0.69 &             0.20            &  150  &   0.41 &     0.06        & 25  &             0.84 &             0.11 \\
    {\em CRISP} &       150 &             0.68 &             0.20            &  250  &   0.41 &     0.06        & 50  &             0.84 &             0.11 \\
    {\em CRISP} &       250 &             0.67 &             0.19            &  500  &   0.41 &     0.06        & 100 &             0.84 &             0.11 \\
    \bottomrule &
    \end{tabular*}
    \caption{Uncertainty estimation results (average over 3 random seeds) for different values of M for our {\em CRISP} method. \vspace{-0.75cm}}
    \label{tab:resutls}
\end{table}

\vspace{-0.6cm}
\section{von Mises-Fisher kernel}

To define the kernel used at the end of section 2, we find the maximum likelihood parameters for the mean direction, $\mu$, and the concentration parameter, $\kappa$, describing the vMF of all the samples $\bar{H}$. The mean direction of the distribution is defined with 
\begin{equation}
    \vec{h}_m = \frac{1}{N} \sum_i^N \vec{h}_i \;\;\;\mbox{ and }\;\;\;     \mu = \frac{\vec{h}_m}{{r}_m}
\end{equation}
where ${r}_m = ||\vec{h}_m||$. We estimate the concentration parameter for a \mbox{$D_h$-dimensional} space using the following equation \cite{kappa_estimate}:

\begin{equation}
    \kappa = \frac{r_m(D_h - r_m^2)}{1-r_m^2}.
\end{equation}
We use these values to define the kernel bandwidth following Taylor's method~\cite{TAYLOR_kernal_bandwith}: 
\begin{equation}
    b = \kappa^{-\frac{1}{2}} (\frac{40\sqrt{\pi}}{N})^{\frac{1}{5}}.
\end{equation}
%
%

\section{Edge uncertainty}

{\em Edge} is a simple morphological edge-detection method.  Let $\delta$ and $\xi$ be dilation and erosion operations, $f^n(\cdot)$ a set of $n$ successive applications of a morphological operator $f$ ($f^0(\cdot)$ is no operation) and $I$ the predicted segmentation map, the {\em Edge} uncertainty map is computed as

\begin{equation}
    U_{edge} = \sum_{n=1}^{5} \Big( \xi^{n}(I) - \xi^{n-1}(I) + \delta^{n}(I) - \delta^{n-1}(I)\Big) \cdot \Big( 1 - n/5 \Big)
\end{equation}

\clearpage
\section{Supplementary results}






\begin{figure}
\centering
\begin{subfigure}{.5\textwidth}
  \centering
  \includegraphics[width=1\textwidth]{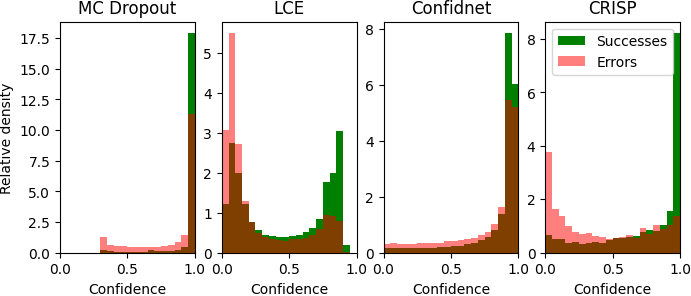}
  \caption{}
  \label{sublable1}
\end{subfigure}%
\begin{subfigure}{.5\textwidth}
  \centering
  \includegraphics[width=1\textwidth]{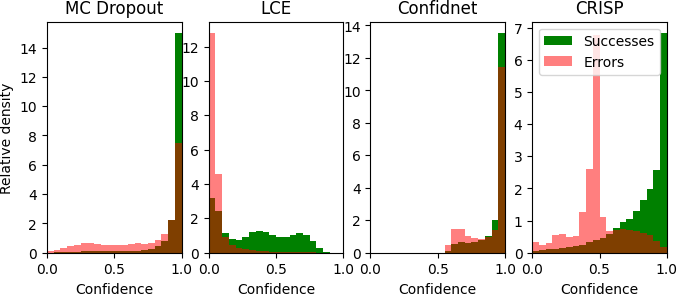}
  \caption{}
  \label{sublable2}
\end{subfigure}
\caption{Histograms on the CAMUS (\ref{sublable1}) and JSRT datasets. (\ref{sublable2}).}
\label{fig:test}
\end{figure}

\vspace{-0.5cm}

\begin{figure*}[h!]
\centering
\includegraphics[width=\textwidth]{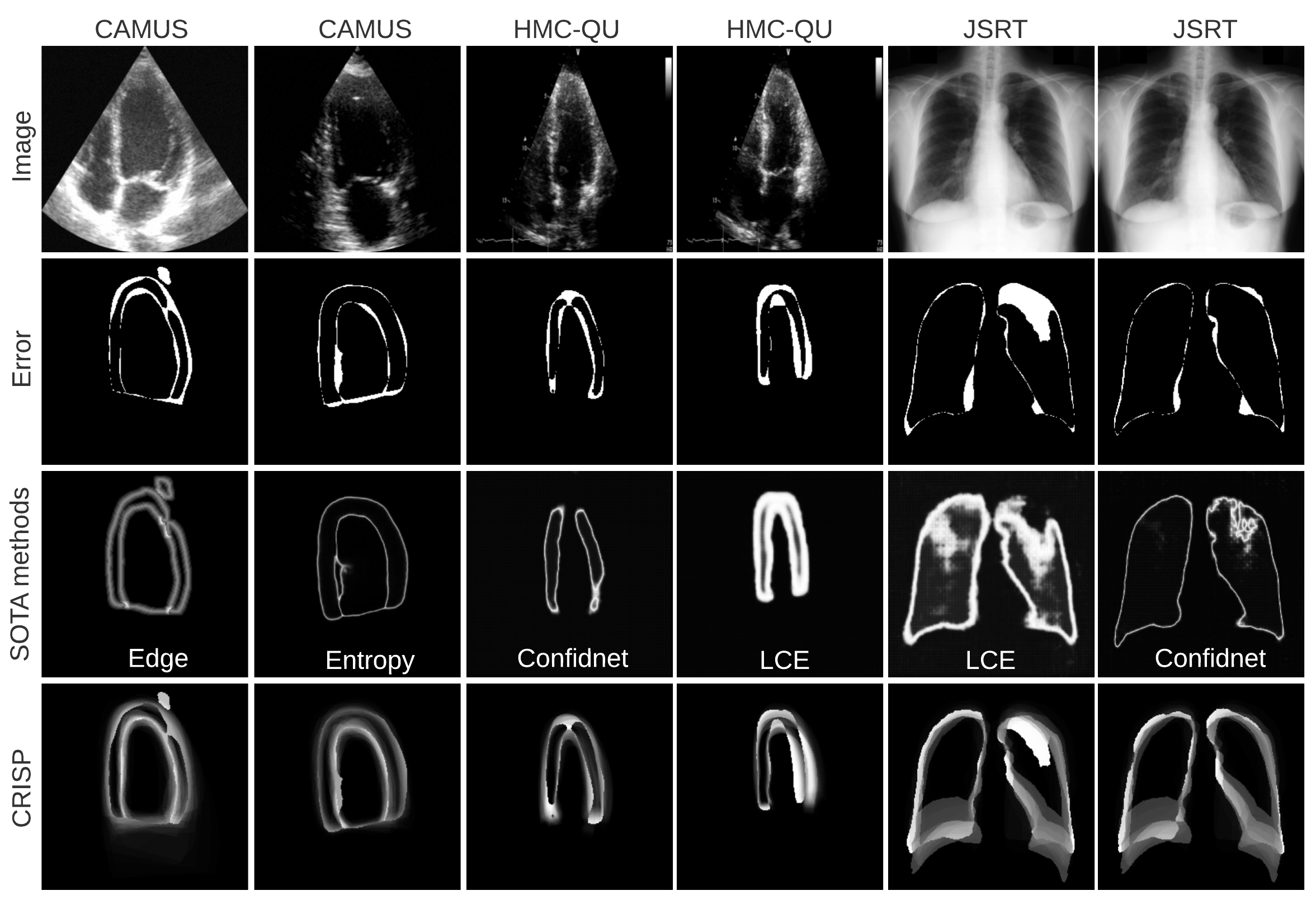}
\caption{Supplementary samples for Fig. 2. \vspace{-0.5cm}}
\label{fig:samples}
\end{figure*}

\bibliographystyle{splncs04}
\bibliography{mybibliography}